\documentclass[referee]{aa} 
\usepackage{graphicx}
\usepackage{txfonts}
\begin{document}
  \title{The bottleneck of the CNO burning and the age of the Globular Clusters}


  \author{G. Imbriani\inst{1,2,3},
          H. Costantini\inst{4}, 
          A. Formicola\inst{5,6}, 
          D. Bemmerer\inst{7}, 
          R. Bonetti\inst{8}, 
          C. Broggini\inst{9}, 
          P. Corvisiero\inst{4}, 
          J. Cruz\inst{10}, 
          Z. F\"ul\"op\inst{11}, 
          G. Gervino\inst{12}, 
          A. Guglielmetti\inst{8}, 
          C. Gustavino\inst{6}, 
          G. Gy\"urky\inst{11}, 
          A.P. Jesus\inst{10}, 
          M. Junker\inst{6}, 
          A. Lemut\inst{4}, 
          R. Menegazzo\inst{9}, 
          P. Prati\inst{4}, 
          V. Roca\inst{2,3}, 
          C. Rolfs\inst{9}, 
          M. Romano\inst{2,3}, 
          C. Rossi Alvarez\inst{9}, 
          F. Sch\"umann\inst{5}, 
          E. Somorjai\inst{11}, 
          O. Straniero\inst{1,2},
          F. Strieder\inst{5}, 
          F. Terrasi\inst{2,13}, 
          H.P. Trautvetter\inst{5}, 
          A. Vomiero\inst{14}, 
          S. Zavatarelli\inst{4}
          }

   \offprints{G. Imbriani, imbriani@na.infn.it}

   \institute{INAF-Osservatorio Astronomico di Collurania, Italy
         \and
             INFN-sezione di Napoli, Italy  
         \and    
             Dipartimento di Scienze Fisiche, Universit\'a Federico II di Napoli, Italy
         \and 
             Universit\'a di Genova, Dipartimento di Fisica and INFN, Genova, Italy
        \and      
             Ruhr Universit\"at Bochum, Bochum, Germany
        \and
             INFN Laboratori Nazionali del Gran Sasso, Assergi, Italy
         \and 
             Institut f\"ur Atomare Physik und Fachdidaktik, Technische Universit\"at Berlin
         \and 
             Universit\'a di Milano, Istituto di Fisica Generale Applicata and INFN, Milano, Italy
         \and 
             INFN-sezione di Padova, Italy
         \and 
             Centro de Fisica Nuclear da Universidade de Lisboa, Lisboa, Portugal
        \and
             Atomki, Debrecen, Hungary
        \and
             Universit\'a di Torino, Dipartimento di Fisica Sperimentale and INFN, Torino, Italy
        \and
             Seconda Universit\'a di Napoli, Dipartimento di Scienze Ambientali, Caserta, and INFN, Napoli, Italy
        \and
             Universit\'a di Padova, Dipartimento di Fisica, Padova and INFN Legnaro, Italy
             }

   \date{Received ......, .....; accepted ......., .....}

\abstract{The transition between the Main Sequence and the Red Giant Branch in low mass stars
is powered by the onset of the CNO burning, whose bottleneck is
the $^{14}$N(p,$\gamma)^{15}$O. The LUNA collaboration has recently
improved the low energy measurements of the cross section of this key reaction.
We analyse the impact of the revised reaction rate on the estimate of the Globular
Clusters ages, as derived from the turnoff luminosity. 
We found that the age of the oldest Globulars should be increased by about 0.7-1 Gyr 
with respect to the current estimates.
  
\keywords{stars: evolution -- stars: Population II --
                Nuclear reactions -- Galaxies: globular clusters: general -- Galaxy: formation --
                cosmology: distance scale}
   }
\titlerunning{The bottleneck of the CNO and globular cluster age}
\authorrunning{G. Imbriani, et al}
   \maketitle

\section{Introduction}

Globular Clusters (GCs) represent the oldest resolved stellar populations.
Their age practically coincides with the time elapsed since the epoch  
of the formation of the first stars in the Universe 
and provides an independent check of the reliability 
of standard (and non standard) cosmological models.
Moreover, the age spread in the GC system is a primary indicator of the 
time scale of the halo formation. 
Among the various methods to date stellar Clusters,
the most reliable and widely adopted is that based on the
measurement of the luminosity of the turnoff 
(i.e. the bluest point on the main sequence).
This dating technique requires the knowledge of the Cluster distance,
the light extinction along the line of sight and the chemical composition
(see \cite{Gra.2} for an exhaustive analysis of the present state of the art).
In addition, a reliable 
theoretical calibration of the
turnoff luminosity-age relation (TOL-A) is needed. 
This relies on our knowledge of the physical 
processes of energy
generation (e.g. nuclear reactions) and transport (e.g. opacity) 
taking place in H burning low mass stars. 
An adequate description of the thermodynamics of stellar
matter is also required. Finally we have to consider any mechanism capable to 
modify the internal chemical 
stratification (once again nuclear reactions, convective mixing, 
rotational induced mixing, microscopic
diffusion or levitation induced by radiation pressure).
\cite{chab} discuss the influence of various theoretical uncertainties
 on the calibration of the turnoff luminosity-age relation and conclude
that the total uncertainty due to the theory may be confined within 0.5 Gyr.   

This paper is devoted to the evaluation of the impact
on the theoretical calibration of the Globular Clusters ages
of the improved determination of the rate of the 
key reaction $^{14}$N(p,$\gamma)^{15}$O, as obtained by the LUNA collaboration 
(\cite{stra.20}).
Since thermonuclear reactions are responsible for chemical modifications 
occurring in stellar interiors
and supply most of the energy irradiated from the stellar surface, the estimated stellar 
lifetime depends on accurate measurements of their rates.
In the last few years, many efforts have been spent in improving these measurements at energy
as close as possible to the Gamow peak, namely the relevant energies at which 
nuclear reactions take place in stars. 
This is a mandatory requirement for the calibration of stellar ages.  

The main sequence stars presently observed in Globular Clusters have masses 
smaller than that of the
Sun. As in the Sun, these low mass stars burn H in the center, mainly through the pp
chain. However, towards the end of their life, when the central hydrogen mass fraction becomes
smaller than about 0.1, the nuclear energy released
by the H burning becomes insufficient and the stellar core must contract to extract
some energy from its own gravitational field. Then, the central temperature (and the density) 
increases and the H burning switches from the pp chain to the more efficient CNO cycle.
Thus, the departure from the Main Sequence is powered by the CNO cycle,
whose bottleneck is the $^{14}$N(p,$\gamma$)$^{15}$O reaction. The luminosity of the turnoff
depends on the rate of this key reaction: the larger the rate, the fainter the turnoff.
On the contrary, the total lifetime is only marginally affected by a change in the CNO,
because it is mainly determined by the rate of the $^1$H(p,$e^+,\nu_e)^2$H.
As a consequence, an increase of the CNO rate would imply fainter turnoff points, 
for a given age, or younger ages, for a given turnoff luminosity (see also \cite{chab98}).
Note that an equivalent effect
may be caused by the enhancement of the CNO abundances (\cite{rood}, \cite{stra.6}).

In the next section we remind the new measurements of the stellar cross section of
the $^{14}$N(p,$\gamma$)$^{15}$O reaction. Then, in section 3 we present the 
revised turnoff luminosity-age relation (TOL-A). 
We show that this revision leads to systematically larger estimates of the age
of the Globular Clusters. Implications for cosmology are 
briefly discussed in the conclusive section.

\section{The updated $^{14}$N(p,$\gamma$)$^{15}$O reaction rate}

The minimum energy explored in nuclear physics laboratories before LUNA was $\sim$240 keV, 
which is well above the range of interest for the stellar CNO burning (${\sim}$20-80 keV).
Therefore, the reaction rate used in stellar model computations is largely extrapolated, 
in a region where the resonant structure of the $^{15}$O compound nucleus is particularly complex.
The rates reported by the popular compilations (\cite{CF88}, CF88, and \cite{NACRE}, NACRE), 
which are based 
on the cross section measurements obtained by \cite{stra.15} are very similar. In particular, 
the astrophysical factor 
at zero energy is S$(0)=3.2\pm0.8$~keV~b (NACRE). 
The main contributions to S(0) come from the transitions to the 
groundstate in $^{15}$O and to the subthreshold state at E$_{cm}=-504$~keV.
It is the existence of this subthreshold resonance that makes the extrapolation
very uncertain.  
Recently \cite{stra.14}, re-analyzing the Schr\"{o}der's experimental data by means of 
a R-matrix model,
report a significant lower S(0), namely $1.77\pm0.20$~keV~b. 
The main discrepancy concerns the contribution of
the captures to the $^{15}$O groundstate, which has been found 19 times smaller than 
the value quoted by Schr\"{o}der and adopted by NACRE and CF88. We emphasize that
the large discrepancy among different analyses based on the same data set is a clear demonstration
of the inadequacy of the low energy extrapolation for this reaction.

The LUNA collaboration\footnote{LUNA is an acronym of the Laboratory for 
Underground Nuclear Astrophysics, operating at the LNGS of Assergi, Italy.}
has significantly improved the low energy measurements of this reaction rate (\cite{stra.20}).
We used a 400 keV facility (\cite{stra.19}), which is particularly well suited
when reaction $\gamma$-ray lines up to $\simeq~7.5$ MeV
have to be measured with very low intensities. Cosmic background is 
strongly suppressed by the mountain shielding and low intrinsic activity 
detectors are employed. The explored energy window ranges from 390 keV down to 135 keV, i.e., 
significantly closer to the astrophysical relevant energy than any previous experiment.
The fit of the new data by means of a R-matrix model leads to
$S(0) = 1.7~\pm~0.1~(stat)~\pm~0.2~(sys)$~keV~b. In the following we use this result to 
revise the calibration of the turnoff luminosity-age relation.

\begin{figure}
\centering
\includegraphics[width=13cm]{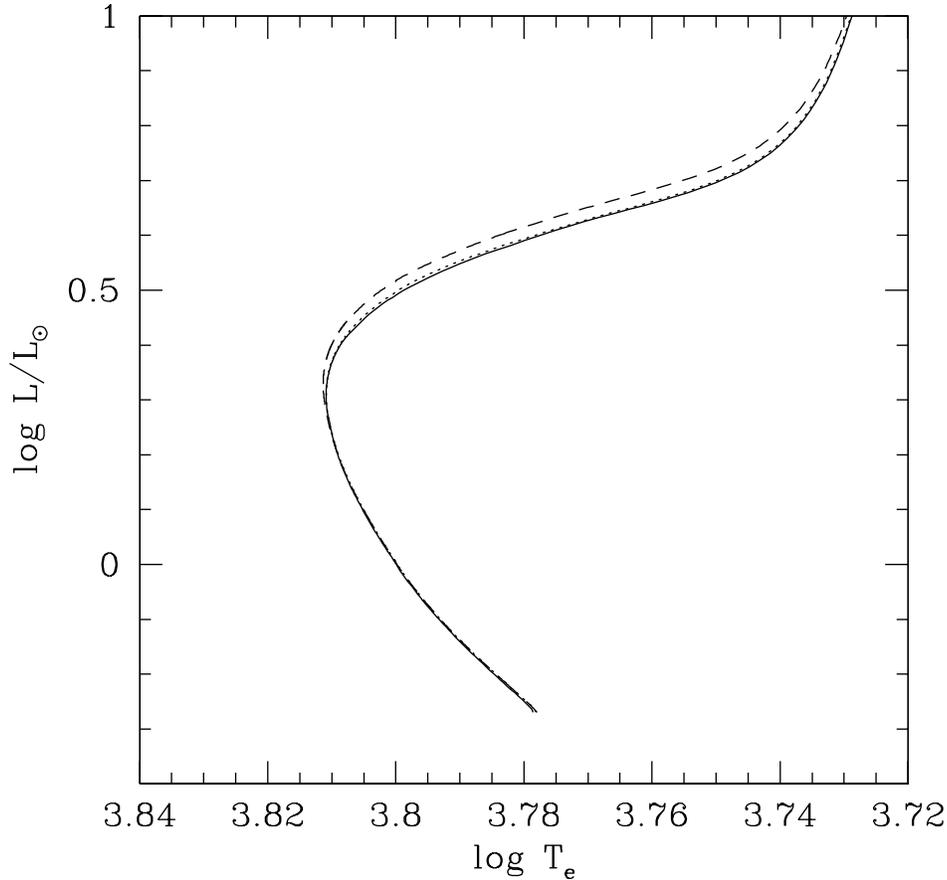}
\caption{Comparison between evolutionary sequences obtained 
with different rates of $^{14}$N(p,$\gamma)^{15}$O: NACRE (dotted line), CF88 (solid line)
and LUNA (dashed line). The stellar mass is 0.8 M$_\odot$ and the metallicity is Z=0.0003}
\label{Fig1}
\end{figure}
   \begin{figure*}
   \centering
   \includegraphics[width=13cm]{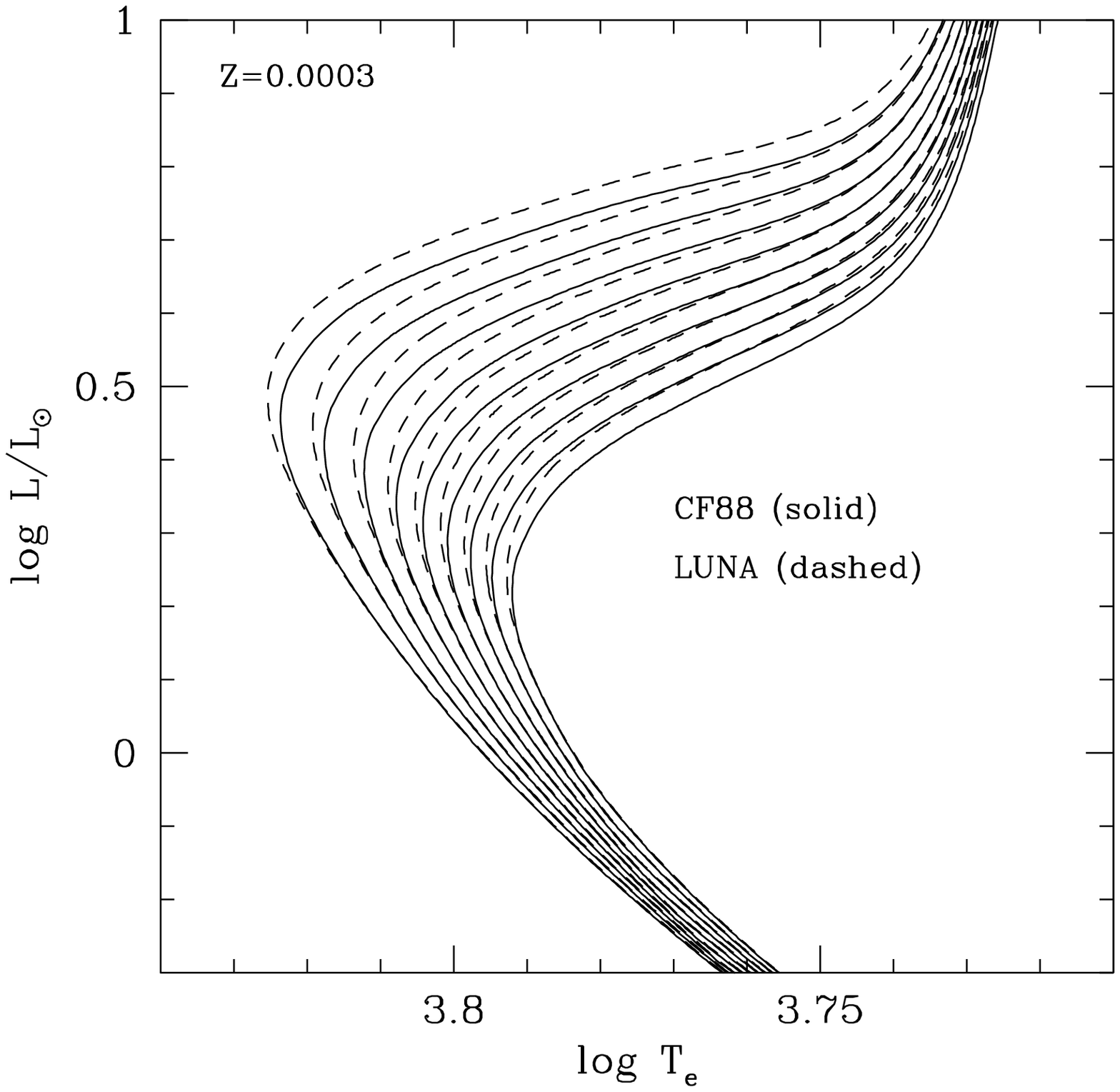}
      \caption{Isochrones for Globular Clusters obtained with different rates of the 
$^{14}$N(p,$\gamma)^{15}$O reaction: CF88 (solid) and LUNA (dashed). The brightest
isochrone (of each set) is the youngest (10 Gyr), while the fainter is the oldest (18 Gyr).
The age step between two adjacent isochrones is 1 Gyr. The metallicity is Z=0.0003 
(or [M/H]=-1.82).}
         \label{Fig2}
   \end{figure*}
   \begin{figure*}
   \centering
   \includegraphics[width=13cm]{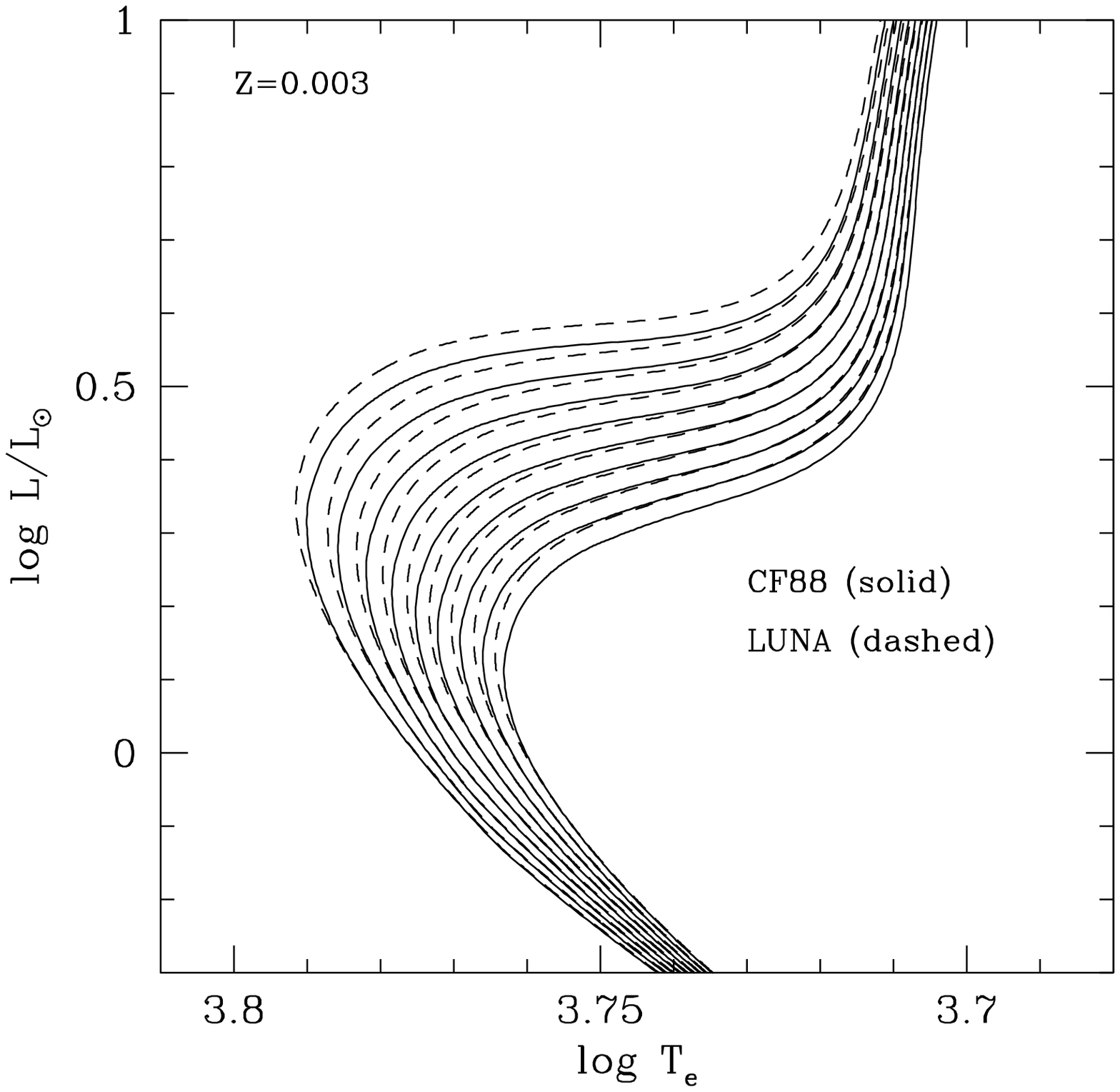}
      \caption{Isochrones for Globular Clusters obtained with different rates of the 
$^{14}$N(p,$\gamma)^{15}$O reaction: CF88 (solid) and LUNA (dashed). The brightest
isochrone (of each set) is the youngest (10 Gyr), while the fainter is the oldest (18 Gyr).
The age step between two adjacent isochrones is 1 Gyr. 
The metallicity is Z=0.003 ([M/H]=-0.52) in the right panel. }
         \label{Fig3}
   \end{figure*}

\section{Globular Clusters ages}

New stellar models have been computed with the same code described in 
\cite{stra.10}, but updating the rate of 
$^{14}$N(p,$\gamma$)$^{15}$O.
We recall that this code includes an improved equation 
of state (which properly takes into account the degree of degeneracy of the electrons and the
electrostatic interactions), the most recent compilation of opacity for stellar interiors 
(\cite{stra.25, stra.24}) and microscopic diffusion
(\cite{stra.27}). Figure 1 shows
an example of the evolutionary track obtained by adopting different rates for the 
$^{14}$N(p,$\gamma$)$^{15}$O reaction. 
The tracks obtained with the CF88 rate practically coincide with the one obtained with
the NACRE rate. On the contrary, the turnoff and the subgiant branch of the sequences
obtained by adopting the new rate are substantially brighter.   

Isochrones have been computed for two sets of stellar models, 
the first based on the old CF88 rate and the second based on the revised LUNA rate.
We have explored the whole
range of chemical composition covered by the galactic GC system. 
In particular, the mass fraction of metals (the
metallicity) has been varied between $Z=0.0001$ and $Z=0.006$,
 which corresponds to [M/H]$=-2.3$ and 
[M/H]$=-0.5$ \footnote{standard spectroscopic notation.}. Some examples 
of the comparison between old and new isochrones are shown in Figure 2 and 3. 
As expected, the lower rate of $^{14}$N(p,$\gamma$)$^{15}$O leads to 
brighter and bluer turnoff points (for a given age).
When a given turnoff luminosity is considered, 
the revised isochrones imply systematically older ages, namely between
0.7 and 1 Gyr.

To compare our isochrones to the available photometric studies of globular cluster stars, we have
transformed luminosities and effective temperatures into magnitudes and colors
 by means of model atmospheres provided by Castelli et al. (1997).
The accuracy of the new isochrones 
in reproducing the morphology of the observed color-magnitude diagrams,
has been checked by selecting two clusters which are representative
 of the oldest component of the galactic halo.
The first test is illustrated in Figure 4. Isochrones for $Z=0.0003$ ([M/H]$=-1.8$) 
and age 13, 14 and 15 Gyr are over imposed
to the color magnitude diagram of NGC 6397. A similar test, 
but for NGC 5904 (M5), is reported in 
Figure 5, where the isochrones have $Z=0.001$ ([M/H]$=-1.3$). 
Photometric data are from the ground-based database published by
Rosenberg et al. (2000). 
In both cases, the new isochrones  match the overall color-magnitude diagram well at 14 Gyr, 
with a bona fide uncertainty of $\pm 1$ Gyr. Similar results were obtained by \cite{stra.10}
with the old (CF88)
isochrones, but, in that case, the best reproduction of the observed diagrams required 13 Gyr
(see their Figure 11).

The following
relation for the Globular Cluster age ($t_9$ in Gyr), as a function of the V magnitude 
of the turnoff point ($M_V^{TO}$) and the metallicity ([M/H]), has been derived.

\begin{eqnarray}
\log t_9 &=& -1.0146-0.2731 {\rm [M/H]}+0.03032 {\rm [M/H]} M_V^{TO}-0.00058 ({\rm [M/H]} M_V^{TO})^2+0.4801 M_V^{TO} \nonumber
\end{eqnarray}

The standard deviation of the estimated age is $\Delta \log t_9=0.005$.
This relation can be used
for ages ranging between 10 and 18 Gyr and [M/H] between $-2.3$ and $-0.5$.

   \begin{figure*}
   \centering
   \includegraphics[width=13cm]{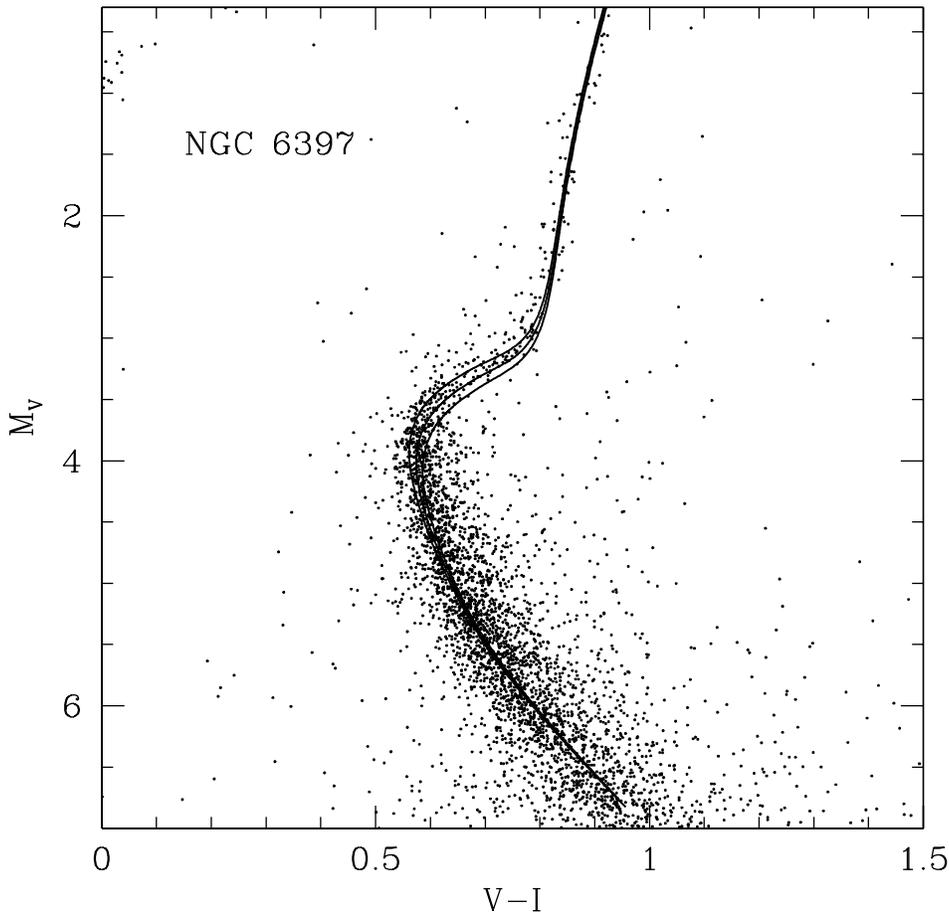}
      \caption{Test to the CMD of the metal-poor cluster NGC 6397. 
The new isochrones with 13, 14 and 15 Gyr are reported. Their metallicity is
Z=0.0003 ([M/H]=-1.8). We adopt (m-M)$_V$=12.58 and E(B-V)=0.18.
The data are from Rosenberg et al. (2000). }
         \label{Fig4}
   \end{figure*}
   \begin{figure*}
   \centering
   \includegraphics[width=13cm]{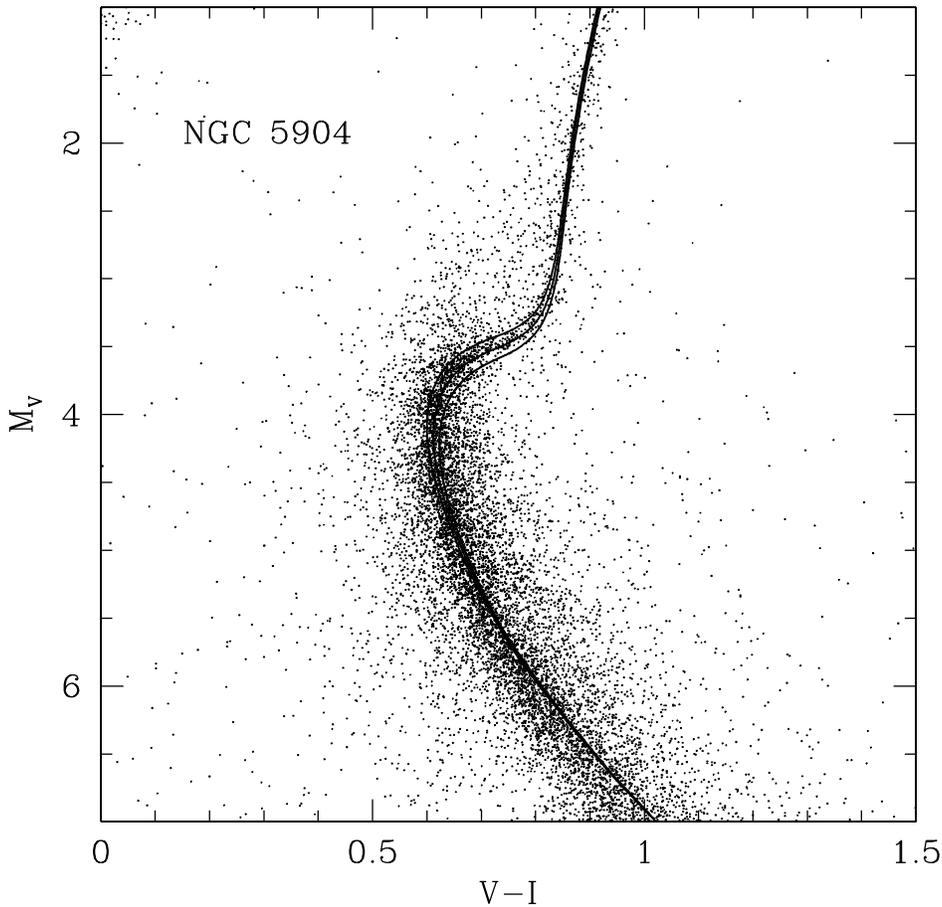}
      \caption{Test to the CMD of the intermediate metallicity cluster NGC 5904. 
The new isochrones with 13, 14 and 15 Gyr are reported. Their metallicity is
Z=0.001 ([M/H]=-1). We adopt (m-M)$_V$=14.41 and E(B-V)=0.02.
The data are from Rosenberg et al. (2000). }
         \label{Fig5}
   \end{figure*}

\section{Implications for cosmology}

The recent developments of accurate measurements of the fundamental 
cosmological parameters allow us to derive a very precise age of the Universe:
$t_0=13.7 \pm 0.2$ Gyr (\cite{stra.28}). 
This result has been obtained in the framework of a $\Lambda$CDM model
and it is based on the measures of three fundamental parameters:
$H_0$, whose best determination has been obtained by the Key HST Project (\cite{stra.17}),
$\Omega$, measured by WMAP (\cite{stra.28}),
and the ratio $\Omega_M/\Omega_{\Lambda}$, constrained by 
 the observation of type Ia supernovae in high redshift Galaxies
(\cite{stra.18}, \cite {smith}).
The galaxy clustering shape measurements also constrain $\Omega_M$
(\cite{percival}).
It is obvious that any systematic uncertainty
affecting just one of these experiments would imply a revision of this estimate of the 
 age of the Universe. For example, it has been argued that the light curve of an
SNe Ia might depend on the chemical composition and/or the mass of the progenitor star.
 In this case, the commonly assumed similarity 
between nearby and high redshift SNe Ia, in spite of the different 
stellar populations of their host Galaxies, could have induced a systematic error in the 
evaluation of $\Omega_M/\Omega_{\Lambda}$ (see e.g. \cite{stra.26})
    
In this context, an independent determination of the age of the Universe, 
may (or may not) confirm the
standard cosmological model that emerges from the experimental cosmology.
At present, the most reliable dating technique 
is the one based on the TOL-A relation for the oldest stellar systems of the Milky Way,
the Globular Clusters.
The standard cosmological model also predicts that the H reionization,
which should
coincide with the epoch of the first star formation, occurred between 100 and 400 Myr after
the Big Bang (95\% CL, \cite{stra.28}). Such a delay must be also considered.

An exhaustive comparison between stellar 
and cosmological ages requires a detailed statistical analysis taking into account 
all sources of errors (experimental and theoretical).
This is beyond the purpose of the present paper and will be presented elsewhere.
 Let us limit our discussion to the expected implication 
of the revised 
$^{14}$N(p,$\gamma$)$^{15}$O reaction rate.
We have shown that the revised ages of the Globular Clusters are older, about 0.7-1 Gyr, than 
those previously claimed. It is worth to note that,  in the framework of the 
$\Lambda$CDM model, 
an equivalent increase of $t_0$ might be obtained by reducing 
$H_0$ ($\sim$5\%) or $\Omega_M$ ($\sim$8\%). These variations are, in any case, within the
experimental errors.

Gratton et al (2003), by means of the TOL-A relation derived  from models taking into account
 the effect of 
microscopic diffusion, but computed adopting the old $^{14}$N(p,$\gamma$)$^{15}$O,
conclude that the
age of the oldest Galactic Clusters is $13.4$ Gyr ($\pm 0.8$ random, $\pm 0.6$ systematic).
When the age increment implied by the revision of the  $^{14}$N(p,$\gamma$)$^{15}$O
 is barely added, the best fit to the age of the oldest GCs should increase above 14 Gyr.
This revised lower limit of the age of the Universe 
strengthens the need of a 
positive cosmological constant. In the case of a flat Universe ($\Omega=1$) and
assuming $H_0=0.72$ Km s$^{-1}$ Mpc$^{-1}$ (\cite{stra.17}), a Universe older than
14 Gyr would imply $\Omega_M<0.22$ or, adopting an uncertainty of 1.4 Gyr in $t_0$ 
(\cite{Gra.2}), $\Omega_M<0.35$.
Note that this upper limit for the matter density is independent of the SNe Ia observations.  
Alternatively, by coupling our result with that of the high redshift SNe Ia, 
we may relax the assumption on the geometry of the Universe to derive a stringent constraint for 
the Hubble constant. Indeed, taking  $H_0 t_0=0.96~\pm~0.04$ (\cite{tonry}),
the present lower limit for $t_0$ would imply $H_O<67$ Km s$^{-1}$ Mpc$^{-1}$ 
(or $H_0<74$ within 1 $\sigma$ in $t_0$), in good agreement with
$72\pm8$ obtained by the Key HST Project (\cite{stra.17}).

\begin{acknowledgements}
We are indebted with the referee (A. Weiss), who found a mistake in the first version 
of the paper. This work has been partially supported by the Italian grant COFIN 2001 and
by FEDER (POCTI/FNU/41097/2001).
\end{acknowledgements}


\begin{thebibliography}{}
\bibitem[Alexander \& Ferguson, 1994]{stra.24} Alexander, D. R. \& Ferguson, J. W., 1994 ApJ, 437, 879
\bibitem[Angulo et al, 1999]{NACRE} Angulo, C. et al, 1999, Nucl. Phys. A 656, 3, (NACRE)
\bibitem[Angulo \& Descouvement (2001)]{stra.14} Angulo, C., \& Descouvemont, P., 2001, Nucl. Phys. A,  690, 755
\bibitem[Castelli et al (1997)]{castelli} Castelli, F., et al, 1997, AA, 318, 841
\bibitem[Caughlan \& Fowler, 1988]{CF88} Caughlan, G.R. \&  Fowler, W.A., 1988, A.D.N.D.T., 40, 283, (CF88) 
\bibitem[Chaboyer et al (1996)]{chab} Chaboyer, B., Demarque, P., Kernan P.J., \& Kraus, L.M., 1996, Science, 271, 975
\bibitem[Chaboyer et al, 1998]{chab98} Chaboyer, B., Demarque, P., Kernan P.J., \& Kraus, L.M., 1998, ApJ, 494, 96
\bibitem[Dom\'\i nguez, H\"oflich \& Straniero, 2001]{stra.26} Dom\'\i nguez, I., H\"oflich, P., \& Straniero, O., 2001, ApJ, 557, 279
\bibitem[Formicola et al, 2003a]{stra.19} Formicola, A., Imbriani, G., Junker, M., et al, 2003a, Nucl.Instr.Meth. A 507, 609 
\bibitem[Formicola et al, 2003b]{stra.20} Formicola, A., Imbriani, G., Costantini, H., et al, 2003b, nucl-ex/0312015 (submitted to Phys. Lett.B) (LUNA) 
\bibitem[Freedman et al, 2001]{stra.17} Freedman, et al, 2001, APJ, 553,47
\bibitem[Gratton et al, 2003] {Gra.2} Gratton et al, 2003, AA, 408, 529
\bibitem[Iglesias, Rogers \& Wilson, 1992]{stra.25} Iglesias, C. A., Rogers, F. J., \& Wilson, B. G. 1992, ApJ, 397, 717 (OPAL)
\bibitem[Percival et al, 2001]{percival} Percival,W. J., et al. 2001, MNRAS, 327, 1297
\bibitem[Perlmutter et al, 1999]{stra.18} Perlmutter, et al, 1999, APJ, 517, 565
\bibitem[Rood, 1981]{rood} Rood, R.T., 1981, in Physical Process in Red Giants, ed. I. Iben Jr., \& A. Renzini (Dordrecht: Reidel), 51
\bibitem[Rosenberg et al, 2000]{rose} Rosenberg, A., Piotto, G., Saviane, I. \& Aparicio, A., 200, AA, 144, 5
\bibitem[Schr\"{o}der et al (1987)]{stra.15} Schr\"{o}der et al, 1987, Nucl. Phys. A,  467, 240
\bibitem[Salaris, Chieffi \& Straniero, 1993]{stra.6} Salaris, M., Chieffi, A. \& Straniero, O., 1993, ApJ, 414, 580
\bibitem[Smith et al, 1998]{smith} Schmidt, B. P., Suntzeff, N. B., Phillips, M. M. et al, 1998, ApJ  507, 46
\bibitem[Spergel et al, 2003]{stra.28} Spergel, D.N. et al, 2003, ApJS 148, 175S
\bibitem[Straniero, Chieffi \& Limongi (1997)]{stra.10} Straniero, O., Chieffi, A. \& Limongi, M., 1997, ApJS, 490, 425
\bibitem[Thoul, Bahcall \& Loub, 1994]{stra.27} Thoul, A.A., Bahcall, J.N., Loub, A., 1994, ApJ, 421, 828
\bibitem[Tonry et al, 2003]{tonry} Tonry, J.L., Schmidt, B.P., Barris, B.,  et al, 2003, ApJ, 594, 1
\end{thebibliography}
\end{document}